\documentclass{ws-ijqi}
\usepackage{psfrag,bm}

\renewcommand{\trimmarks}{\relax}
\renewcommand{\draftnote}{\relax}

\def\currenttime{\hour=\time \divide\hour by 60 \number\hour:%
  \multiply\hour by 60 \minute=\time \global\advance\minute by -\hour%
  \ifnum\minute<10 0\number\minute\else\number\minute\fi}

\newcommand{\scsz}[1]{\mbox{\scriptsize$#1$}}

\newcommand{\JournalTitle}[1]{\textit{#1}\ }

\newcommand{\PRA}{\JournalTitle{Phys.\ Rev.\ A}}
\newcommand{\PRL}{\JournalTitle{Phys.\ Rev.\ Lett.}}

\begin{document}

\markboth{S. M. Assad, J. Suzuki, B.-G. Englert}%
{Raw-data attacks in quantum cryptography with partial tomography}

\title{%
\uppercase{Raw-Data Attacks in Quantum Cryptography with Partial Tomography}}

\author{\uppercase{Syed M. Assad, Jun Suzuki, Berthold-Georg Englert}}
\address{Department of Physics, %
National University of Singapore, Singapore 117542\\
assad@singmail.com, physj@nus.edu.sg, phyebg@nus.edu.sg}

\maketitle

\begin{history}
\received{22 September 2006}
\end{history}

\begin{abstract}
We consider a variant of the BB84 protocol for quantum cryptography, the
prototype of tomographically incomplete protocols, where the key is generated
by one-way communication rather than the usual two-way communication.
Our analysis, backed by numerical evidence, establishes thresholds for
eavesdropping attacks on the raw data and on the generated key at quantum bit
error rates of 10\% and 6.15\%, respectively. 
Both thresholds are lower than the threshold for unconditional security in the
standard BB84 protocol.
\end{abstract}

\bigskip\bigskip


\section{Introduction}\label{sec:intro}
Quantum cryptography deals with all aspects, both theoretical and
experimental, of schemes, or ``protocols,'' for quantum key distribution.
When implementing 
a quantum key distribution protocol, the two parties --- Alice and Bob ---
exploit the laws of quantum mechanics to gain a string of perfectly
secure key bits for subsequent one-time-pad encryption of a private message. 

The first such protocol, the celebrated BB84 protocol, was proposed by Bennett
and Brassard in 1984.\cite{Bennett1984:1} In this brief report, we study
raw-data attacks on this protocol in an ``Ekert setting,'' where 
eavesdropper Eve is given the privilege of sending entangled qubits to
Alice and Bob. This setting is equivalent to the traditional scenario
where Alice sends qubits to Bob, and Eve performs a cloning attack on the
qubits in transmission.

Alice and Bob concede to the fact that there will always be noise in
their channel, but they only accept unbiased noise. Now, owing to the
incomplete-tomographic nature of this protocol, Eve is free to send
a family of states, with each of them appearing the same to Alice and Bob.
For them, this state is only parameterized by one parameter, the amount
of noise they see. We ask this question: which state should Eve send to
Alice and Bob such that she maximizes her accessible information relative 
to Alice?

We restrict the discussion to the case where the state
Eve sends is symmetric between Alice and Bob. For all such states,
a von Neumann type measurement that satisfies the necessary conditions
for an optimal POVM --- in the sense of maximizing the mutual
information --- is given. But even in this restricted case we cannot prove
that the POVM found is globally optimal. As usual\cite{BookChapter} we can
verify that we found a local optimum, but any claim on global optimality must
rely on a thorough numerical search with negative outcome.

A plausible, self-suggesting answer to that question would be for Eve to send
a state to Alice and Bob such that it has maximal entropy. But it turns out
that this is not always her best choice. 
Rather, when attacking the raw data Eve should
send the state with the largest degree of separability, 
or smallest concurrence.
From the information that is then accessible to Eve, we deduce the
security threshold for the raw-data attacks.

\section{Partial tomography}\label{sec:PartTomo}
The scenario we shall be discussing is as follows: Alice and Bob are
promised a sequence of singlets from Eve, an entangled qubit pair
provider of dubious reliability. Due to practical imperfections in
their channel, they compromise with an unbiased-noise state%
\footnote{In practice, Alice and Bob would not expect
to receive such a state with perfectly unbiased noise. However, they can
include a controlled effective source of noise in their apparatus, by
post-processing the measured data, and so bring any
state they receive to the unbiased-noise state.}
\begin{equation}\label{eq:unbiased-noise}
\rho_\mathrm{AB}
=(1-\epsilon)\left|\phi_1\right\rangle \left\langle \phi_1\right|
+\frac{\epsilon}{4}\,.
\end{equation}
Here $\left|\phi_1\right\rangle$ is the singlet state, and
$0\leq\epsilon\leq1$ characterizes the amount of noise in the
channel; the ``quantum bit error rate'' that is used to quantify the noise in
the standard BB84 protocol equals $\frac{1}{2}\epsilon$.
Alice and Bob independently perform measurements in two complementary bases,
the $x$ and $z$ basis, each with equal probability. The decision of whether to
carry out the measurement in the $x$ or $z$ basis is made at random.

To check the reliability of the source they received, Alice and Bob
sacrifice a fraction of their qubits to verify that the
joint probability table of their measurements is consistent with
(\ref{eq:unbiased-noise}), i.e., they check that their joint probability
table looks like Table~\ref{cap:AB prob table}.
The quantum version\cite{RenatoThesis} of the de Finetti theorem ensures that
the qubit pairs received from the source behave like statistically independent
pairs.
Therefore, the joint probabilities of Table~\ref{cap:AB prob table} tell
Alice and Bob the statistical properties of those pairs as far as measurements
in the $x$ and $z$ bases are concerned, but there is no information about the
$y$ bases.
Accordingly, the ``state tomography'' performed by Alice and Bob in this
manner is incomplete, or \emph{partial}.

\begin{table}[t]
\tbl{\label{cap:AB prob table}Joint probability table between
Alice and Bob who communicate with each other to check that the
statistical properties of their measurement results are consistent 
with this table.}
{\begin{tabular}{ccccc}\toprule
 Bob&
\multicolumn{4}{c}{Alice}\\
&
 $z+$&
 $z-$&
 $x+$&
 $x-$\\%\tabularnewline
\colrule
$z+$&
 $\displaystyle\frac{\epsilon}{16}$&
 $\displaystyle\frac{2-\epsilon}{16}$&
 $\displaystyle\frac{1}{16}$&
 $\displaystyle\frac{1}{16}$\\[2.5ex]
$z-$&
 $\displaystyle\frac{2-\epsilon}{16}$&
 $\displaystyle\frac{\epsilon}{16}$&
 $\displaystyle\frac{1}{16}$&
 $\displaystyle\frac{1}{16}$\\[2.5ex]
$x+$&
 $\displaystyle\frac{1}{16}$&
 $\displaystyle\frac{1}{16}$&
 $\displaystyle\frac{\epsilon}{16}$&
 $\displaystyle\frac{2-\epsilon}{16}$\\[2.5ex]
$x-$&
 $\displaystyle\frac{1}{16}$&
 $\displaystyle\frac{1}{16}$&
 $\displaystyle\frac{2-\epsilon}{16}$&
 $\displaystyle\frac{\epsilon}{16}$\\
\botrule
\end{tabular}}
\end{table}

Alice and Bob, being paranoid, assume that the noise 
in (\ref{eq:unbiased-noise})
is an artifact of Eve's eavesdropping on their communication.
So, if the probability table they obtain is skewed, or if they find
that the noise level $\epsilon$ is too large, Alice and Bob abandon
their communication. Otherwise they will use the correlations in their
data to establish a cryptographic key for one-time-pad encryption.

By means of the partial tomography, Alice and Bob establish the values of eight
independent parameters of their joint two-qubit state. 
Explicitly, if we write the two-qubit state between Alice and Bob as
\begin{equation}\label{eq:rab}
\rho_\mathrm{AB}
=\frac{1}{4}\sum_{j,k=0}^{3}c_{jk}\sigma_{j}\otimes\sigma_{k}\;,
\end{equation}
where $\sigma_{0}=1$ and $\sigma_{j}$ are the three Pauli matrices,
the tomographic constraint of consistency with (\ref{eq:unbiased-noise})
imposes eight constraints on the coefficients 
$c_{jk}=\langle\sigma_j\otimes\sigma_k\rangle$,
\begin{eqnarray}\label{eq:cons1}
&c_{01}=c_{03}=c_{10}=c_{30}=c_{13}=c_{31}=0\;\textrm{ and }&
\nonumber\\ &
c_{11}=c_{33}=-(1-\epsilon)\;. &
\end{eqnarray}
With $c_{00}=1$ for normalization, there remain seven parameters that are
inaccessible to Alice and Bob who, therefore, can ascertain 
the state they receive only partially. 
Hence we speak of \emph{partial tomography}. 
The remaining seven parameters are independent and are constrained only by the
positivity of $\rho_\mathrm{AB}$.

This is in contrast to, say, the ``six-state protocol'' or the ``minimal
qubit protocol'' where Alice and Bob perform full tomography.%
\cite{Liang2003:1,SingProt}
There they can check all $15$ parameters of their
joint two-qubit state, and tomography uniquely characterizes the state. 
In addition to (\ref{eq:cons1}), Alice and Bob then have the luxury of 
insisting on
\begin{eqnarray}\label{eq:c02}
&c_{02}=c_{20}=c_{12}=c_{21}=c_{23}=c_{32}=0&
\nonumber\\&\textrm{and}\quad
c_{22} = -(1-\epsilon)
\end{eqnarray}
as well.

In the partially tomographic BB84 protocol, these parameters are hidden from
Alice and Bob. A whole family of distinct states appears equivalent to
them. They would be wise to assume that Eve uses her freedom
to manipulate these hidden parameters to her full advantage.

As the scheme in the BB84 protocol goes, Alice and Bob would reveal
publicly the bases of their independent measurements. After this
announcement, qubit pairs measured in mismatched bases are discarded,
while the qubit pairs in matched bases give them a string of sifted
data with stronger correlations. In the absence of noise, these sifted
data would have perfect correlations, and the resulting key 
is guaranteed to be secure. 
However, in the presence of noise, a shorter but still perfectly secure key
can still be distilled by means of classical error correcting codes and privacy
amplification.

But Alice and Bob can just as well device a protocol
that uses the raw data itself; that is, they exploit the correlations available
directly from Table~\ref{cap:AB prob table} to distill the same amount
of secure key bits. Alice and Bob then need not announce their measurement
bases.  In the subsequent analysis, it is about this raw data that 
eavesdropper Eve wishes to learn. This is equivalent to Eve attempting
to eavesdrop on the sifted data if she does not have the means to
store her ancilla qubits until after Alice and Bob will have announced
their choice of bases.

According to the  Csisz\'ar--K\"orner theorem\cite{CK} of classical
information theory, this amount of distillable secure key is measured by the
difference in the Shannon's mutual information between Alice and Bob and
between Alice and Eve. 
The mutual information between Alice and Bob is
\begin{equation}\label{eq:IAB}
I_\mathrm{AB}(\epsilon)=\frac{1}{2}\Phi\left(1-\epsilon\right)\,,  
\end{equation}
where
\begin{equation}
\Phi\left(x\right)=\frac{1}{2}\bigl[\left(1-x\right)\log\left(1-x\right)
+\left(1+x\right)\log\left(1+x\right)\bigr]\,.  
\end{equation}

\section{Constraints on Eve}\label{sec:Eve}
Eve creates an entangled four-qubit state
\begin{equation} \label{eq:psiABE}
\left|\Psi_\mathrm{ABE}\right\rangle =\sum_{j=1}^{4}\left|\phi_{j}\right\rangle
\left|E_{j}\right\rangle\,,
\end{equation}
where the unnormalized kets $\left|E_{j}\right\rangle $ are the
four two-qubit states of her ancilla which record the outcomes of Alice and
Bob's measurements, and 
\begin{eqnarray}\label{eq:Bell}
\left.
\begin{array}{l}
\left|\phi_{1}\right\rangle\\
\left|\phi_{2}\right\rangle
\end{array}
\right\}
& = & \Bigl( \left|z+\right\rangle \left|z-\right\rangle \mp 
\left|z-\right\rangle \left|z+\right\rangle \Bigr) \frac{1}{\sqrt{2}} 
\nonumber\\ 
\textrm{and}\quad\left.
\begin{array}{l}
\left|\phi_{3}\right\rangle\\
\left|\phi_{4}\right\rangle
\end{array}
\right\}
& = & \Bigl( \left|z+\right\rangle \left|z+\right\rangle 
\pm \left|z-\right\rangle \left|z-\right\rangle \Bigr) \frac{1}{\sqrt{2}}
\end{eqnarray}
are the four Bell states, which we use as the basis states for the qubit pair
received by Alice and Bob. 
The two-qubit state obtained by performing a partial trace over
Eve's ancilla,
\begin{equation}
\rho_\mathrm{AB}=\textrm{Tr}_\mathrm{E}\Bigl\{
\left|\Psi_\mathrm{ABE}\right\rangle 
\left\langle \Psi_\mathrm{ABE}\right|\Bigr\}\,,  
\end{equation}
is what Eve sends to Alice and Bob. The geometry of Eve's ancilla states are
fully determined by $\rho_\mathrm{AB}$, the state Eve chooses, and the free
choice of basis we used for Alice and Bob in writing (\ref{eq:psiABE}). 
With $\rho_\mathrm{AB}$ in the form of (\ref{eq:rab}), Eve is thus constrained
by the values of $c_{jk}$ in (\ref{eq:cons1}).

\section{Raw-data attacks}\label{sec:attacks}
Eve's task is to maximize the mutual information between Alice and
herself.  For every state $\rho_\mathrm{AB}$ that Eve chooses to send, she has
a corresponding optimal POVM that maximizes her mutual
information.  Hence hers is a double optimization problem:
first, she has to find the best POVM and, second, she has to choose the most
advantageous values for the seven adjustable coefficients that do not appear
in (\ref{eq:cons1}).
The optimal POVM depends, of course, on the parameter choice.

The optimization has to be done with respect to Eve's four input states.
They are the ancilla states conditioned on Alice measuring $z\pm$ or
$x\pm$. 
Each of these states has rank two. 

In practice, Eve does not have to make use of ancillas for an attack on the
raw data. 
Once she decides which POVM to use on her ancillas, she can trace out her
subsystem from the state (\ref{eq:psiABE}) conditioned on her intended POVM
outcomes. The remaining ensemble of states would then give the
pre-manufactured states that Eve should send to Alice and Bob.

We restrict our study to the symmetric case of 
$c_{02}=c_{20}=c_{12}=c_{21}=c_{23}=c_{32}=0$, where $\rho_\mathrm{AB}$ is
symmetric under the interchange of Alice and Bob, and all expectation values
$c_{jk}=\langle\sigma_j\otimes\sigma_k\rangle$ vanish if either
$\sigma_j=\sigma_2$ or $\sigma_k=\sigma_2$ but not both.
Although there exist
subspaces outside the symmetric region where the accessed information
equals the accessed information in the symmetric region, numerical
simulations suggest strongly that Eve has no advantage from nonsymmetric
states. 

\begin{figure}[t]
\begin{center}
\includegraphics[bb=181 392 436 631,clip=]{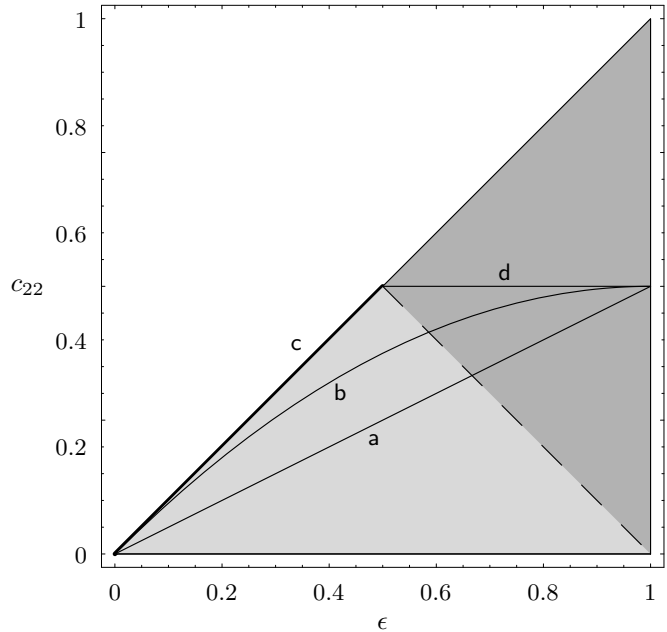}
\caption{\label{c22plot}The shaded area indicates the region of permissible
values of $c_{22}$ so that $\rho_\mathrm{AB}$ is positive. In the darker shaded
region, $\rho_\mathrm{AB}$ is separable. 
The straight line \textsf{a} corresponds to the state
that Eve would be restricted to if she were forced to send the 
state (\ref{eq:unbiased-noise}), which is the case when Alice and Bob 
perform complete tomography. 
For $c_{22}$ values on curve \textsf{b}, Eve sends a maximum-entropy state
to Alice and Bob. The thick line \textsf{c} gives the maximum-separability
state for a fixed $\epsilon< 1/2$. Along line \textsf{d}, where $c_{22}=0$, Eve
can have $1/2$ bits of mutual information with Alice. Since the accessible
information decreases with $\left|c_{22}\right|$, the lines \textsf{c} and
\textsf{d} provide Eve with the largest accessible information.}
\end{center}
\end{figure}

In this symmetric regime, then, Eve's four ancilla states in the basis
specified by (\ref{eq:psiABE}) are mutually orthogonal, 
\begin{equation}
  \langle E_j|E_k\rangle=\delta_{jk}\langle E_j|E_j\rangle\,,
\end{equation}
so that the right-hand side of (\ref{eq:psiABE}) is the Schmidt decomposition
of $|\Psi_\mathrm{ABE}\rangle$,
and $\rho_\mathrm{AB}$ is a weighted sum of projectors on the Bell states
(\ref{eq:Bell}) with the weights given by
\begin{eqnarray}\label{eq:Ejprobs}
 \langle E_1|E_1\rangle&=&\frac{1}{4}(3-2\epsilon-c_{22})\,,\nonumber\\
 \langle E_2|E_2\rangle&=&\langle E_4|E_4\rangle=\frac{1}{4}(1+c_{22})\,,
\nonumber\\
 \langle E_3|E_3\rangle&=&\frac{1}{4}(-1+2\epsilon-c_{22})\,.
\end{eqnarray}
The positivity of $\rho_\mathrm{AB}$ thus requires 
\begin{equation}
  \label{eq:c22range}
              -1\leq c_{22}\leq2\epsilon-1\,,  
\end{equation}
which identifies the shaded area in Figure~\ref{c22plot}. 
 
The reduced ancilla states that are conditioned on Alice getting one of her
four measurement results $z+$, $z-$, $x+$, or $x-$ are then given by
\begin{eqnarray}\label{eq:ancillastates}
  \rho_{z\pm}&=&
\bigl(|E_1\rangle\pm|E_2\rangle\bigr)\bigl(\langle E_1|\pm\langle E_2|\bigr)
+\bigl(|E_3\rangle\pm|E_4\rangle\bigr)\bigl(\langle E_3|\pm\langle E_4|\bigr)
\,,\nonumber\\
  \rho_{x\pm}&=&
\bigl(|E_1\rangle\mp|E_4\rangle\bigr)\bigl(\langle E_1|\mp\langle E_4|\bigr)
+\bigl(|E_2\rangle\pm|E_3\rangle\bigr)\bigl(\langle E_2|\pm\langle E_3|\bigr)
\,,
\end{eqnarray}
each of them occurring with probability $\frac{1}{4}$. 
It is thus Eve's task to discriminate between these states as best as she can,
by a suitable POVM, whereby the figure of merit is the accessed information,
equal to the mutual information between Eve and Alice that results from the
chosen POVM.

\section{POVMs that maximize the mutual information}\label{sec:POVM}
The mutual information that Eve achieves is maximized, at least locally,
by a von Neumann measurement composed of the projectors to the following kets:
\begin{eqnarray}\label{eq:POVM}
\left.
\begin{array}{l}
\left|P_{1}\right\rangle\\
\left|P_{2}\right\rangle
\end{array}
\right\}
& = & 
\left|E_{1}\right\rangle\frac{1}{\sqrt{3-2\epsilon-c_{22}}}
\pm \left|E_{2}\right\rangle\ 
\sqrt{\frac{3-2\epsilon-c_{22}}{1-{c_{22}^{\ }\,\!}^2}} 
\nonumber\\&&
- \left|E_{3}\right\rangle\ \frac{i}{\sqrt{2\epsilon-1-c_{22}}}
\mp \left|E_{4}\right\rangle\ i
\sqrt{\frac{2\epsilon-1-c_{22}}{1-{c_{22}^{\ }\,\!}^2}}
 \nonumber\\\textrm{and}\quad
\left.
\begin{array}{l}
\left|P_{3}\right\rangle\\
\left|P_{4}\right\rangle
\end{array}
\right\}
& = & 
\left|E_{1}\right\rangle  \frac{1}{\sqrt{3-2\epsilon-c_{22}}}
\pm \left|E_{2}\right\rangle\ 
i \sqrt{\frac{2\epsilon-1-c_{22}}{1-{c_{22}^{\ }\,\!}^2}} 
\nonumber\\&&
+ \left|E_{3}\right\rangle\ \frac{i}{\sqrt{2\epsilon-1-c_{22}}}
\mp \left|E_{4}\right\rangle\ 
\sqrt{\frac{3-2\epsilon-c_{22}}{1-{c_{22}^{\ }\,\!}^2}}
\,.
\end{eqnarray}
The mutual information obtained from this POVM is
\begin{equation}\label{eq:mi-ae}
I_\mathrm{AE}=\frac{1}{2}\Phi\left(\sqrt{1-{c_{22}^{\ }\,\!}^{2}}\,\right)\;,
\end{equation} independent of $\epsilon$.

In fact, this POVM is not the only one that attains this mutual
information. 
Since Eve's conditioned ancilla states (\ref{eq:ancillastates}) have real
coefficients in the basis of the $|E_j\rangle$ kets, the complex conjugate of
(\ref{eq:POVM}) gives exactly the same mutual information. 
And so does any convex combination
of these two POVMs, although the members of the so-formed POVM are
no longer of rank one. In particular, an equal-weight combination gives an
optimal POVM with real coefficients. 
Consult Ref.~\refcite{BookChapter} for more details about this matter.

Eve makes use of her freedom to select any $c_{22}$ value within the limits of
(\ref{eq:c22range}) such that $I_\mathrm{AE}$ is largest.
This amounts to choosing the smallest permissible value of $\bm|c_{22}\bm|$.
For $\epsilon\geq\frac{1}{2}$, this is $c_{22}=0$, giving the straight line
(d) in Fig.~\ref{c22plot}; for $\epsilon\leq\frac{1}{2}$ the best choice is
$c_{22}=-(1-2\epsilon)$, which traces out line (c) in Fig.~\ref{c22plot}.

Accordingly, Eve has
\begin{equation}
  \label{eq:IAEopt}
 I_\mathrm{AE}(\epsilon)=\left\{
   \begin{array}{c@{\ \textrm{for}\ }l}
     \frac{1}{2}\Phi\Bigl(2\sqrt{\epsilon(1-\epsilon)}\,\Bigr)
     & 0\leq\epsilon\leq\frac{1}{2}\,,\\[1ex]
 \frac{1}{2} &  \frac{1}{2}\leq\epsilon\leq1\,,
   \end{array}\right.
\end{equation}
after optimizing the value of $c_{22}$.
The comparison with $ I_\mathrm{AB}(\epsilon)$ of (\ref{eq:IAB}) then implies 
that the BB84 protocol is secure under raw-data attacks when
$\epsilon<\frac{1}{5}$, which corresponds to a quantum bit error rate of 10\%.

Figure~\ref{miplot} shows $I_\mathrm{AE}(\epsilon)$ for the $c_{22}$ values
along curves  \textsf{a}, \textsf{b}, and \textsf{c} in Fig.~\ref{c22plot},
for the relevant range of $0\leq\epsilon\leq\frac{1}{2}$.
Also shown is $I_\mathrm{AB}(\epsilon)$ of (\ref{eq:IAB}), which decreases as
$\epsilon$ increases.

\section{Largest degree of separability, smallest concurrence}
\label{sec:DoS-concur}
\begin{figure}[t]
\begin{center}
\includegraphics[bb=180 455 436 635,clip=]{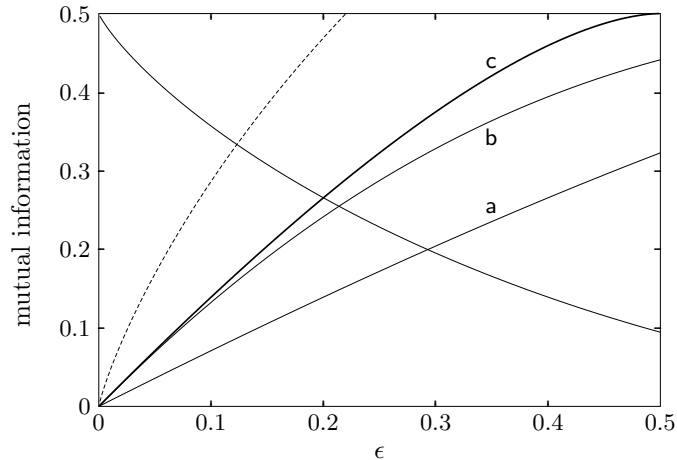}
\caption{\label{miplot}
  The decreasing function shows the mutual information $I_\mathrm{AB}$
  between Alice and Bob as given in (\ref{eq:IAB}). 
  Curve \textsf{a} gives Eve's accessible information (\ref{eq:mi-ae}) when she
  honestly sends the unbiased-noise state (\ref{eq:unbiased-noise}). 
  Eve is restricted to this if Alice and Bob were to perform complete
  tomography on their states. 
  Curve \textsf{b} plots Eve's accessible information when she sends a state
  with maximum entropy, and curve \textsf{c} applies when she sends the
  state with the largest degree of separability, or the smallest concurrence,
  and thus achieves the optimum of (\ref{eq:IAEopt}). 
  The curves \textsf{a}, \textsf{b}, and \textsf{c} intersect Alice and Bob's
  mutual information at $\epsilon=1-\sqrt{1/2}=0.2929$, 
  $\epsilon=1-\sqrt{\sqrt{5/4}-1/2\, }=0.2138$, 
  and $\epsilon=1/5=0.2000$, respectively. --- 
  The dashed curve shows $I_\mathrm{AE}^\mathrm{(HSW)}(\epsilon)$ of
  (\ref{eq:HSWopt}). }
\end{center}
\end{figure}

As is clearly shown by Figs.~\ref{c22plot} and \ref{miplot}, Eve's best choice
for $c_{22}$ does not amount to sending the unbiased-noise state
(\ref{eq:unbiased-noise}) to Alice and Bob, for which ${c_{22}=-(1-\epsilon)}$,
nor the two-qubit state with the largest entropy, for which
${c_{22}=-(1-\epsilon)^2}$.\cite{Englert1994:1}
Rather, Eve sends the state with the largest degree of separability
$\mathcal{S}$, a quantity introduced by Lewenstein and Sanpera,\cite{L+S} and
the smallest value of the Hill--Wootters 
concurrence~$\mathcal{C}$.\cite{Wootters}

For the two-qubit state $\rho_\textrm{AB}$ under consideration, which is
diagonal in the Bell-state basis of (\ref{eq:Bell}) and thus 
``self-transposed'' in the terminology of Ref.~\refcite{EngMet}, one has
\begin{eqnarray}
  \label{eq:S+C}
  \mathcal{S}&=&\min\Bigl\{1,\epsilon+\frac{1}{2}\bigl(1+c_{22}\bigr)\Bigr\}\,,
\nonumber\\
  \mathcal{C}&=&\max\Bigl\{0,\frac{1}{2}\bigl(1-c_{22}\bigr)-\epsilon\Bigr\}\,,
\end{eqnarray}
so that $\mathcal{S}+\mathcal{C}=1$ and maximizing $\mathcal{S}$ is tantamount
to minimizing $\mathcal{C}$.
For other families of states, however, different states may realize the
largest $\mathcal{S}$ value and the smallest $\mathcal{C}$ value.
Since the self-transposed states do not offer a clue, we leave it as a moot
point which of the two quantities is the crucial one.

\section{Maximal entropy}\label{sec:maxEnt}
The POVM of Sec.~\ref{sec:POVM} and the thresholds of Fig.~\ref{miplot}
apply when it is Eve's objective to gain maximal knowledge about Alice's
measurement results, for each qubit pair sent to Alice and Bob.
In other words, Eve is attacking the raw data that have the correlations of
Table~\ref{cap:AB prob table}.

These correlations are turned into a cryptographic key by a suitable
error-correcting code for one-way communication from Alice to Bob.
For this purpose, the measurement results are identified with the letters of
an alphabet --- such as $(z+,z-,x+,x-)\widehat{=}$(A,B,C,D) for Alice
and $(z+,z-,x+,x-)\widehat{=}$(B,A,D,C) for Bob --- and the code words are
sequences of A, B, C, and D. 
Alice chooses at random one of the code words and informs Bob, over a public
channel, which of her measurement results make up the code words (``Listen,
it's qubits 101, 17, 53, 2674, \dots'').
Bob's corresponding measurement results constitute the received word, which he
can then decode to the code word Alice sent. 
The sequence of transmitted code words, or perhaps a single very long code
word, are then the key for the one-time-pad encryption.

Clearly, Eve is much more interested in this key than in the raw data from
which it is generated.
If she has the technical means for storing her ancillas, she will not measure
them until after Alice has publicly announced the qubit pairs that contribute
to the code words. 
Only then will Eve measure the respective ancillas jointly, thereby gaining
more information per qubit pair, possibly as much as the
Holevo--Schumacher--Westmoreland bound\cite{HSW} grants,
\begin{equation}
  \label{eq:HSW}
  I_\mathrm{AE}^\mathrm{(HSW)}
    =S(\rho_\mathrm{E})-\frac{1}{4}\!\!
 \sum_{\alpha=\left\{\!\genfrac{}{}{0pt}{}{\scsz{z\pm}}{\scsz{x\pm}}\right.}
      \!\!S(\rho_\alpha)\,,
\end{equation}
where
\begin{equation}
  \rho_\mathrm{E}=\frac{1}{4}\sum_{\alpha}\rho_\alpha
                 =\sum_{j=1}^4\bigl|E_j\bigr\rangle\bigl\langle E_j\bigr|
\end{equation}
is the over-all ancilla state, and 
$S(\rho)=-\mathrm{tr}\{\rho\log_2\rho\}$ is the von
Neumann entropy in units of bits.
The non-zero eigenvalues of each $\rho_\alpha$ are $1-\frac{1}{2}\epsilon$ and
$\frac{1}{2}\epsilon$, and the eigenvalues of $\rho_\mathrm{E}$ are the
probabilities of (\ref{eq:Ejprobs}), so that $I_\mathrm{AE}^\mathrm{(HSW)}$ is
readily evaluated.

Since $S(\rho_\alpha)$ does not depend on $c_{22}$, the $c_{22}$ value for
which $I_\mathrm{AE}^\mathrm{(HSW)}$ is largest, is the value for which
$\rho_\mathrm{E}$ has maximal entropy.
It is also the $c_{22}$ value for which $\rho_{AB}$ has maximal entropy
because $\rho_\mathrm{E}$ and $\rho_{AB}$ are unitarily equivalent.
As stated above, this happens for $c_{22}=-(1-\epsilon)^2$.
Then
\begin{equation}
  \label{eq:HSWopt}
  I_\mathrm{AE}^\mathrm{(HSW)}(\epsilon)=1-\Phi(1-\epsilon)
               =1-2I_\mathrm{AB}(\epsilon)\,,
\end{equation}
so that the corresponding threshold value for $\epsilon$ is determined by
$I_\mathrm{AB}(\epsilon)=\frac{1}{3}$. 
This gives $\epsilon=0.1230$, or a quantum bit error rate of 6.15\%, as is
illustrated by the dashed curve in Fig.~\ref{miplot}.

\section{Conclusion}
We have considered quantum key distribution from the raw-data correlations of
the BB84 scheme by one-way communication, and have established the noise
thresholds for eavesdropping attacks on the raw data and on the generated key.
Owing to the incomplete state tomography in the BB84 scenario, Eve can choose
the two-qubit state she sends to Alice and Bob from a seven-parameter family.
Our analysis invokes a plausible symmetry conjecture, which is unproven as yet
but backed by numerical evidence, namely that Eve can restrict herself to
sending self-transposed states to Alice and Bob, which have only one free
parameter. 

We find that the noise threshold for the raw-data attack is
$\epsilon=\frac{1}{5}$, which Eve achieves by distributing the two-qubit state
with the smallest concurrence, or the largest degree of separability, to Alice
and Bob.
By contrast, the ultimate attack on the generated key is most powerful when
the two-qubit state with maximal entropy is sent, and the resulting threshold
is at $\epsilon=0.1230$.
The corresponding quantum bit error rates, after the bases matching in the
standard BB84 protocol, are 10\% and 6.15\%, respectively.
Comparison with the accepted threshold for unconditional security
for BB84, about 12.4\%,\cite{BB84-threshold} thus establishes that the key
extraction by two-way communication (bases matching etc.) is advantageous for
Alice and Bob.

\section*{Acknowledgments}
J.~S. and B.-G.~E. wish to thank Hans Briegel for the generous hospitality
extended to them at the Institute for Quantum Optics and Quantum Information
in Innsbruck, where part of this work was done. 
This work is supported by A*STAR Temasek Grant No.~012-104-0040 and 
NUS Grant WBS R144-000-116-101.

\end{document}